# Efficient and accurate analysis of oscillation dynamics for dissipative cavity solitons based on the artificial neural network

**Mao-Lin Wang,**[1, †] **Peng-Xiang Wang,**[1, †] **Gang Xu**[1, *]

[1]*School of Optical and Electronic Information, Huazhong University of Science and Technology, 1037 Luoyu Road, Wuhan, China*
[†]*These authors contributed equally to this work.*
*\*gang_xu@hust.edu.cn*

**As a conventional means to analyze the system mechanism based on partial differential equations (PDE) or nonlinear dynamics, iterative algorithms are computationally intensive. In this framework, the details of oscillating dynamics of cavity solitons are beyond the reach of traditional mathematical analysis. In this work, we demonstrate that this long-standing challenge could be tackled down with the Long Short-Term Memory (LSTM) neural network. We propose the incorporating parameter-fed ports, which are capable of recognizing period-doubling bifurcations of respiratory solitons and quickly predicting nonlinear dynamics of solitons with arbitrary parameter combinations and arbitrary time series lengths. The model predictions capture oscillatory features with a small Root Mean Square Errors (RMSE) = 0.01676 and an absolute error that barely grows with the length of the prediction time. Lugiato-Lefever equation (LLE) based parameter space boundaries for typical oscillatory patterns are plotted at about 120 times the speed relative to the split-step Fourier method (SSFM) and higher resolution.**

Dissipative structure theory is a science that studies the properties of dissipative structures and the laws of their formation, stability and evolution [1],[2], which has been continuously refined and proven over the past century. Different localized structures (LS), represented by dissipative solitons (DS), are studied in a wide range of nonlinear open systems, including chemistry [3], biology [4], nonlinear optics [5], and population dynamics [6]. Through the sustaining exchange of matter and energy with the outside world, when the change of external conditions reaches a certain threshold, the self-organization phenomenon can be generated through the internal action, so that the system can spontaneously transform from the original disordered state to the ordered state in temporal and space domain, and form a new, stable or periodic DS. In recent years, the growing field of ultrafast optics has provided efficient discoveries for dissipation theory and is becoming an excellent platform for the study of DSs in the pumping-dissipation system, including integrated microcavities [7], mode-locked lasers [8] and the coherently-driven passive fiber ring resonators [9].

However, these emerging platforms suffer from excessively rough simulation calculations and overly complex model building. To be specific, due to the limitation of partial differential equation (PDE) based on nonlinear Schrödinger equation in optical fiber, traditional split-step Fourier method (SSFM) must take into account the accuracy and speed of model evolution [10]. SSFM is increasingly difficult to compute for complex and coupled systems [11],[12], especially when exploring the dynamics of unknown respiratory soliton states. In addition, the dynamic behavior in the nonlinear respiratory soliton state is essentially and urgently needed for the study of chaos, spectral analysis and optical fiber communication. In this mission, one proposed a faster, more accurate, data-only (free from complex models) solution – Artificial Neural Network [13].

Over the past few decades, artificial intelligence (AI) technologies have made significant strides, particularly in data-driven modeling and prediction. With advancements in computational capabilities and the widespread use of big data, Recurrent Neural Networks (RNNs) have emerged as a powerful deep learning architecture for dynamic predictions in complex systems. Dynamical systems often exhibit high nonlinearity and complexity, making traditional modeling approaches less efficient. RNNs, with their recursive structure, can capture both long-term and short-term dependencies in the temporal dimension, effectively modeling the time-varying characteristics of system dynamics. For instance, Fang *et al*. employed neural networks for dynamic modeling of stable solitons and soliton molecules in mode-locked lasers [14], while Pu *et al*. utilized RNNs with prior information to predict the dynamics of mode-locked fiber lasers [15]. Unfortunately, conventional RNNs are limited by the disappearance of gradients caused by long-term dependence problems, which means that constant predictions will forget the initial data. This deficiency is particularly significant when predicting respiratory dynamics in lasers or fiber ring resonators, although simulating stable soliton molecules hides this challenge.

In this work, we propose a scheme based on long short-term memory (LSTM) networks to analyze the respiratory DSs of optical systems for the first time. Through a short period of learning (6 hours), the powerful prediction ability accurately matches the existence map of breathers in the coherently-driven passive resonator. In addition, this achievement represents that we can directly find and distinguish different periodic-doubling behaviors according to several key parameters in the equation, avoiding the unessential resource cost caused by enumeration algorithm and aimless exploration of SSFM. Furthermore, the LSTM may meet the needs of PDEs in different fields, not just Lugiato-Lefever equation (LLE) [16] and cubic-quintic Ginzburg-Landau equation (CQGLE) [17],[18], in ultrafast optics. AI technology replaces the complex bifurcation and stability analysis of several PDEs with rapid prediction of steady-state solutions.

To start, we introduce the mean-field LLE model to describe the dynamics of electric field evolution in a nonlinear dielectric cavity. Considering the pump power and detuning as important parameters of the electric field evolution, the dimensionless normalized LLE equation has the following form [16],[19]:

$$\frac{\partial E(t,\tau)}{\partial t} = \left[-1 + i\left(|E(t,\tau)|^2 - \Delta\right) - i\eta\frac{\partial^2}{\partial \tau^2}\right]E(t,\tau) + S \quad (1).$$

Here, $E(t,\tau)$ represents the intracavity electric field, and the terms in the square brackets on the right-hand side of the equation refer to the role of losses, Kerr nonlinearity, and dispersion on the evolution of the optical field, respectively. The key driving parameters, including the uniform plane wave pump power $S$ and the cavity detuning $\Delta$ which play crucial roles for the cavity dynamics. As shown in Fig. 1(a), the coherently pumped passive Kerr resonator may emit stable cavity solitons (CS), homogeneous steady states (HSS), modulation instabilities (MI) patterns, and chaos under specific parameter conditions. In particular, a novel class of special soliton forms of dynamical evolution called *Period-N oscillations* is found both theoretically and experimentally [20]-[22].

When parameters cross the Hopf bifurcation line [red solid line in Fig. 1(b)], stable cavity solitons could be regarded as period-1 oscillations. As the pump power and detuning further increase, more complex multi-periodic oscillations emerge, including temporal chaos, during which the optical field maintains its localization in the fast time scale. With a moderate further increase of the pump power and detuning, due to the disappearance of stable limit cycle attractors in the two-dimensional phase space, the initial state undergoes transient evolution before relaxing to the HSS. In fact, the preliminary parameter space can be plotted through bifurcation and stability analysis [23],[24]. However, the gray area E in Fig. 1(b) covers the region that includes Period-N oscillations (N=1, 2, …), temporal chaos, and spatiotemporal chaos, whose parameter boundaries remain challenging to determine through analytical or numerical methods. Only experimental and manual parameter scanning in simulations have allowed partial access to the characteristics of these regimes [25].

However, within the Kerr cavity, there are multiple states of multi-periodic oscillations, temporal chaos and HSS in the parameter space of the E region shown in Fig. 1(b), resulting in a complex parameter landscape. Additionally, identifying the oscillatory modes at the boundaries where different oscillatory modes switch can be challenging. A common and intuitive method is to observe the limit loop trajectories of their stable evolution in a 2D phase diagram; however, this approach requires extensive simulation data and manual identification. Consequently, delineating the parameter boundaries of different oscillation modes in this region proves to be difficult.

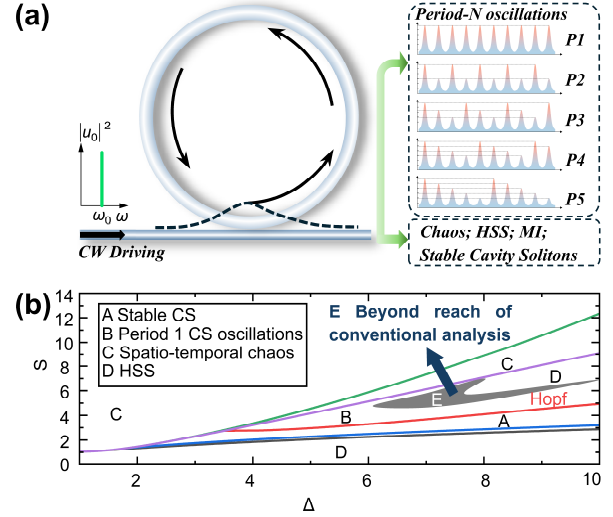

Fig. 1. (a) Output patterns of a representative Kerr resonator, and peak evolution curves of Period-N oscillatory solitons. (b) parameter space of output patterns (defined by detuning vs. pump power).

To address this issue, we propose an artificial neural network model based on Long Short-Term Memory (LSTM) networks. This model incorporates parameter feed ports to enable the modeling of specific equations. Additionally, we utilize the power spectral analysis to process the neural network outputs, facilitating fully automated parameter delimitation and theoretically achieving infinite parameter accuracy.

Our training dataset is derived from simulation data generated using the SSFM. We utilize approximated solutions of the Lugiato-Lefever Equation (LLE) in limiting cases $E = \sqrt{2\Delta}\,\text{sech}\left(\sqrt{\Delta}\tau\right)$ as initial pulse inputs, and a strategy validated by our research and related studies to ensure it does not impact the results of the dynamical evolution [25]. The simulation time step is set to 0.00001 roundtrips to maintain sufficient accuracy in the parameter boundary regions, which are particularly sensitive to value changes, especially near the chaos boundaries. The training dataset consists of 546 samples, with parameters $S$ ranging from 5 to 5.5 and $\Delta$ ranging from 7 to 8. Each sample includes 3000 steps (equivalent to 60 roundtrips, with a down sampling step of 1 step = 0.02 roundtrip) of evolution data. The width of the normalized time window for the fast time scale $\tau_{\text{span}} = 20$, represented by a vector of length 512.

Fig. 2 illustrates the neural network model employed in this study. The data generated using the SSFM is segmented at the slow time scale with a specified window width $W$. It is crucial to select the window width judiciously; an excessively large window may result in inefficient use of computational resources, while a window with insufficient width could hinder the model's ability to capture large-scale evolutionary features. In this letter, the chosen window length constitutes 0.67% of the total data length and approximately 40% of the typical Period-1 oscillation, thereby balancing both efficiency and accuracy. The data and labels are formatted as follows:

$$\begin{cases} data: I(i,\tau), I(i+1,\tau)\cdots I(i+W-1,\tau) \\ label: I(i+W,\tau) \end{cases} \quad (2)$$

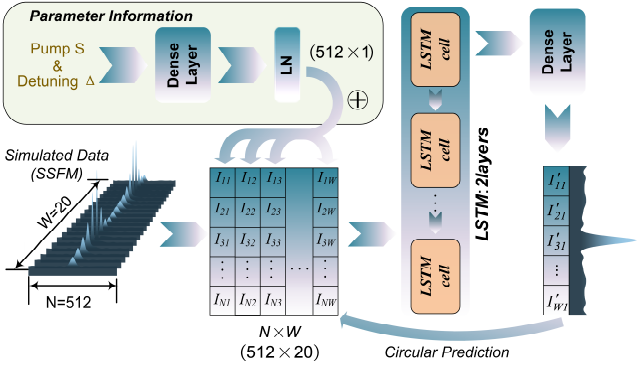

Fig. 2. The pipeline LSTM-based neural network model with parameters feeding.

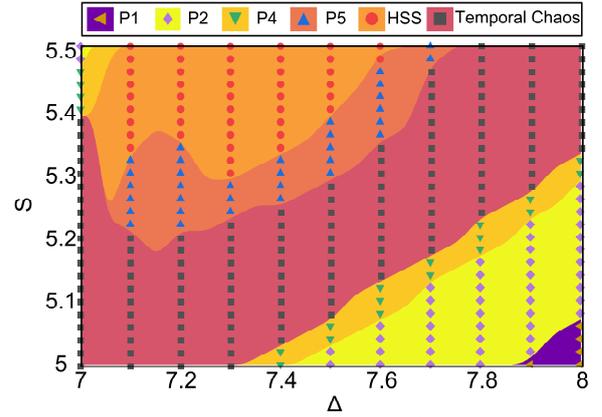

Fig. 3. Colored regions: predicted different output pattern regions (Parameter resolution=0.02). Dots: simulation validation by uniformly sampling 286 sets of parameters.

The windowed data are input into the model, along with pump power and detuning specified as feeding parameters. The inclusion of these a priori parameters enhance the specificity of the model. The final model outputs the predicted data at the next slow time and the model loss function is defined by Root Mean Square Errors (RMSE) with the following function:

$$\ell_{RMSE} = \sqrt{\frac{\sum_{r,c,m}(I_m - \tilde{I}_m)^2}{R \cdot C \cdot M}} \quad (3)$$

In this expression, $I_m$, $\tilde{I}_m$ represents the actual and predicted light field intensity distribution of the m$th$ sample, respectively. And $R$, $C$, $M$ represents the number of rows, columns and total number of samples of the two-dimensional light field intensity distribution. In this example, the dataset includes 546 sets of light field evolution distributions under the parameter sets. After a random shuffle, 491 sets are used as the training set, while the remaining samples serve as the validation set. In the LSTM layer we used a two-layer LSTM superposition to ensure sufficient modeling power and avoid overfitting. The Adam optimizer is used to train 200 epochs and a dynamically decreasing learning rate strategy is used to improve the training stability.

With this idea, we use the initial set of 20 steps of simulation data and the neural network can generate the complete evolution of optical field intensity through a closed-loop circular prediction approach. The soliton peak evolution curve is then extracted from the generated data, followed by power spectrum analysis utilizing the Kaiser window function to reduce the spectral leakage of the FFT to improve the frequency resolution. This method enables the automatic identification of pulse oscillation characteristics and facilitates fully automated parameter region delineation by establishing an appropriate threshold value of characteristic spectrum peak. Our method accelerates the practical approach of manually identifying oscillation behavior through simulation-drawn 2D phase portraits, reducing the average time from 1.99 s/roundtrip to $1.65 \times 10^{-2}$ s/roundtrip.

With this method, automated identification of the oscillation states across various parameter regions is achieved, as shown in Fig. 3. While the figure illustrates a spline curve with a parameter resolution of 0.2, the trained model theoretically allows for the generation of curves with infinite precision. We validated the classification accuracy with 286 uniformly sampled instances, achieving a prediction accuracy rate of 98.60%. In comparison to the traditional SSFM, our approach delivers prediction speeds that are approximately 120 times faster.

It is essential to recognize that accurate identification of Period-N oscillations is heavily dependent on a substantial amount of intensity data. This data can be readily obtained through the predictive capabilities of a trained neural network that is able to predict beyond the training scope. Otherwise, for long-period oscillatory behaviors, such as the P5 oscillation, 3000 steps may contain fewer than 10 complete stable cycles. As a result, the data from numerical simulations may not provide enough Period-N oscillatory cycles to accurately determine the oscillatory behavior from the power spectrum. In our work, the fully automated prediction and mapping of the parameter space matches very well with the point-by-point test results of existing related work [23][25]. The model exhibits strong generalization capabilities, enabling it to perform effectively even with previously unseen data from entirely unfamiliar parameter regions. It generates evolution data that is 2.67 times the slow time range of the data included in the training set which is a feat not accomplished in prior dynamics modeling studies. For the parameter sets, training utilized only 546 sets of parameter data, while in prediction, we forecasted 1326 sets of parameter states, theoretically allowing for an infinitely high density of parameter selection. In Fig. 4, we present the corresponding simulated and predicted evolutions for two sets of parameter states, selecting the two most complex examples of P5 oscillations and collapses to HSS as representative cases, with respective RMSE of 0.00455 and 0.00220 within the first 1000 steps. The comparison of the peak evolution curves illustrates that the model generalizes exceptionally well, predicting 5000 steps (or even more) beyond the training scope. Even though closed-loop prediction implies that errors accumulate, no clear upward trend is seen in the absolute error curves in Fig. 4(b, d).

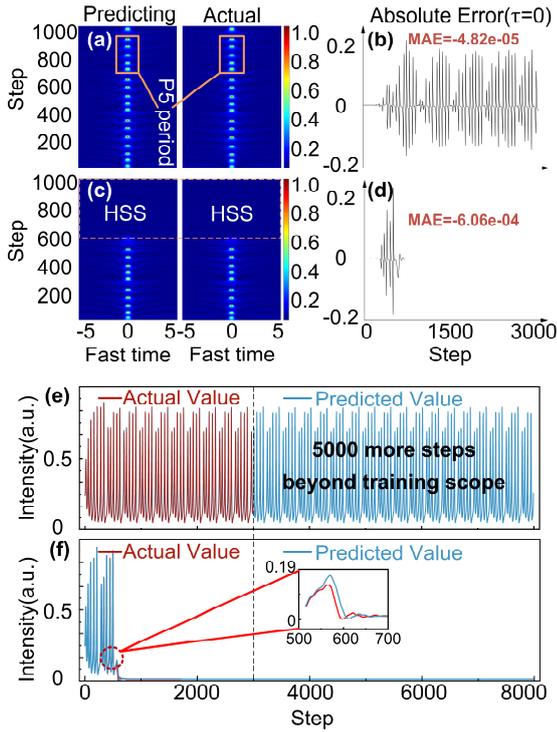

Fig. 4. (a, b, e) $\Delta = 7.15$, $S = 5.2$: Predictions vs. actual evolution of P5 oscillations; Absolute error evolution at $\tau = 0$; Peak evolution curves. (c, d, e) $\Delta = 7.4$, $S = 5.5$: Predictions vs. actual evolution of soliton collapses to HSS; Absolute error evolution at $\tau = 0$; Peak evolution curves. MAE, Mean Absolute Error.

It should be noted that for different parameters, the initial evolution of the windowed data with a length of 20 steps is largely similar, and the model is still able to extract the oscillatory features and chaotic or collapse trends in it and accurately display them after thousands of steps. Meanwhile, for the time-domain chaotic evolution, it is possible that it always remains localized at fast time scales and chaotically evolved at slow time scales under the action of chaotic attractors. It is also possible to undergo a period of transient time-domain chaotic evolution under the perturbation of the cavity soliton stable manifold of saddle-node (caused by the SN bifurcation generating a saddle solution) before collapsing to the stable HSS attractor, and the transient chaotic process follows an evolution similar to that of the chaotic attractor [26]. Even a part of the collapse behavior occurs after 3000 steps, and the model can reproduce this dynamical phenomenon.

The model was trained for approximately 6 hours under CUDA (Compute Unified Device Architecture) acceleration, enabling it to model the dynamics of slow-time processes of arbitrary length at about 120 times the speed with any parameters. The simulation was conducted using MATLAB R2024a on an AMD Ryzen 9 7940HX CPU. The neural network training took place in a Python environment utilizing the PyTorch library, on a cloud server equipped with an Intel(R) Xeon(R) Platinum 8260 CPU @ 2.30GHz and an NVIDIA 3090 graphics card.

In conclusion, this paper provides a paving contribution to the steady-state solution identification of DSs in complex systems using LLE as an example. The model is sufficiently robust to capture the oscillatory features embedded in the equations for specific parameters, and can therefore predict lengths of time well beyond the training scopes, and the absolute errors do not show an upward trend. From a broader perspective, our work represents a purely data-driven artificial neural network framework. This means it can be transferred to any other complex equations at virtually no additional cost. We believe that the combination of neural networks and mathematical analysis is a powerful tool to accurately and efficiently explore the physical mechanisms of much more complex systems, such as coupled parity-time symmetry system [27]-[29], cascaded microcavities [30],[31], and multimode fibers [32],[33].

**Funding.** The National Natural Science Foundation of China (62275097).

**Disclosures**.

**Data availability.** Data underlying the results presented in this paper are not publicly available at this time but may be obtained from the authors upon reasonable request.